\documentclass[10pt,english,10pt, conference, letterpaper]{IEEEtran}
\usepackage[T1]{fontenc}
\usepackage[latin9]{inputenc}
\usepackage{color}
\usepackage{float}
\usepackage{units}
\usepackage{amsthm}
\usepackage{amsmath}
\usepackage{amssymb}
\usepackage{graphicx}

\makeatletter

\floatstyle{ruled}
\newfloat{algorithm}{tbp}{loa}
\providecommand{\algorithmname}{Algorithm}
\floatname{algorithm}{\protect\algorithmname}

\floatstyle{ruled}
\newfloat{algorithm}{tbp}{loa}
\floatname{algorithm}{Algorithm}
\usepackage[noend]{algorithmic}
\newcommand{\forbody}[1]{ #1 \ENDFOR}
\newcommand{\ifbody}[1]{ #1  \ENDIF}

\theoremstyle{plain}
\newtheorem{thm}{\protect\theoremname}
\theoremstyle{plain}
\newtheorem{lem}[thm]{\protect\lemmaname}

\usepackage{bbm}

\@ifundefined{showcaptionsetup}{}{%
 \PassOptionsToPackage{caption=false}{subfig}}
\usepackage{subfig}
\makeatother

\usepackage{babel}
\providecommand{\lemmaname}{Lemma}
\providecommand{\theoremname}{Theorem}

\begin{document}

\title{In-Order Delivery Delay of Transport Layer Coding}

\author{Jason Cloud$^{*}$, Douglas Leith$^{\dagger}$, and Muriel M\'{e}dard$^{*}$\\
$^{*}$ Massachusetts Institute of Technology, Cambridge, MA 02139,
USA\\
$^{\dagger}$ Hamilton Institute, National University of Ireland Maynooth,
Ireland\\
Email: jcloud@mit.edu, doug.leith@nuim.ie, medard@mit.edu}
\maketitle
\begin{abstract}
A large number of streaming applications use reliable transport protocols
such as TCP to deliver content over the Internet. However, head-of-line
blocking due to packet loss recovery can often result in unwanted
behavior and poor application layer performance. Transport layer coding
can help mitigate this issue by helping to recover from lost packets
without waiting for retransmissions. We consider the use of an on-line
network code that inserts coded packets at strategic locations within
the underlying packet stream. If retransmissions are necessary, additional
coding packets are transmitted to ensure the receiver's ability to
decode. An analysis of this scheme is provided that helps determine
both the expected in-order packet delivery delay and its variance.
Numerical results are then used to determine when and how many coded
packets should be inserted into the packet stream, in addition to
determining the trade-offs between reducing the in-order delay and
the achievable rate. The analytical results are finally compared with
experimental results to provide insight into how to minimize the delay
of existing transport layer protocols.

\vspace{-5pt}
\end{abstract}

\section{Introduction}

Reliable transport protocols are used in a variety of settings to
provide data transport for time sensitive applications. In fact, video
streaming services such as Netflix and YouTube, which both use TCP,
account for the majority of fixed and mobile traffic in both North
America and Europe \cite{_sandvine_????}. In fixed, wireline networks
where the packet erasure rate is low, the quality of user experience
(QoE) for these services is usually satisfactory. However, the growing
trend towards wireless networks, especially at the network edge, is
increasing non-congestion related packet erasures within the network.
This can result in degraded TCP performance and unacceptable QoE for
time sensitive applications. While TCP congestion control throttling
is a major factor in the degraded performance, head-of-line blocking
when recovering from packet losses is another. This paper will focus
on the latter by applying coding techniques to overcome lost packets
and reduce head-of-line blocking issues so that overall in-order packet
delay is minimized.

These head-of-line blocking issues result from using techniques like
selective repeat automatic-repeat-request (SR-ARQ), which is used
in most reliable transport protocols (e.g., TCP). While it helps to
ensure high efficiency, one problem with SR-ARQ is that packet recovery
due to a loss can take on the order of a round-trip time ($RTT$)
or more \cite{xia_analysis_2003}. When the $RTT$ (or more precisely
the bandwidth-delay product ($BDP$)) is very small and feedback is
close to being instantaneous, SR-ARQ provides near optimal in-order
packet delay. Unfortunately, feedback is often delayed and only contains
a partial map of the receiver's knowledge. This can have major implications
for applications that require reliable delivery with constraints on
the time between the transmission and in-order delivery of a packet.
As a result, we are forced to look at alternatives to SR-ARQ.

This paper will explore the use of a systematic random linear network
code (RLNC), in conjunction with a coded generalization of SR-ARQ,
to help reduce the time needed to recover from losses. The scheme
considered first adds redundancy to the original data stream by injecting
coded packets at key locations to help overcome potential losses.
This has the benefit of reducing the number of retransmissions and,
consequently, the delay. However, correlated losses or incorrect information
about the network can result in the receiver's inability to decode.
Therefore, feedback and coded retransmission of data is also considered. 

The following sections will provide the answers to two questions about
the proposed scheme: when should redundant packets be inserted into
the original packet stream to minimize in-order packet delay; and
how much redundancy should be added to meet a user's requested QoE.
These answers will be provided through an analysis of the in-order
delivery delay as a function of the coding window size and redundancy.
We will then use numerical results to help determine the cost (in
terms of rate) of reducing the delay and as a tool to help determine
the appropriate coding window size for a given network path/link.
While an in-depth comparison of our scheme with others is not within
the scope of this paper, we will use SR-ARQ as a baseline to help
show the benefits of coding at the transport layer. 

The remainder of the paper is organized as follows. Section \ref{sec:Related-Work}
provides an overview of the related work in the area of transport
layer coding and coding for reducing delay. Section \ref{sec:Network-Model}
describes the coding algorithm and system model used throughout the
paper. Section \ref{sec:Preliminaries} provides the tools needed
to analyze the proposed scheme; and an analysis of the first two moments
of the in-order delay are provided in Sections \ref{sec:Expected-In-Order-Delivery}
and \ref{sec:In-Order-Delivery-Delay}. Furthermore, the throughput
efficiency is derived in Section \ref{sec:Efficiency} to help determine
the cost of coding. Numerical results are finally presented in Section
\ref{sec:Numerical-Results} and we conclude in Section \ref{sec:Conclusion}.\vspace{-6pt}

\section{Related Work\label{sec:Related-Work}}

A resurgence of interest in coding at the transport layer has taken
place to help overcome TCP's poor performance in wireless networks.
Sundararajan et. al. \cite{sundararajan_network_2011} first proposed
TCP with Network Coding (TCP/NC). They insert a coding shim between
the TCP and IP layers that introduces redundancy into the network
in order to spoof TCP into believing the network is error-free. Loss-Tolerant
TCP (LT-TCP) \cite{subramanian_hybrid_2007,tickoo_lt-tcp:_2005,ganguly_loss-tolerant_2012}
is another approach using Reed-Solomon (RS) codes and explicit congestion
notification (ECN) to overcome random packet erasures and improve
performance. In addition, Coded TCP (CTCP) \cite{kim_congestion_2014}
uses RLNC \cite{ho_random_2006} and a modified additive-increase,
multiplicative decrease (AIMD) algorithm for maintaining high throughput
in high packet erasure networks. While these proposals have shown
coding can help increase throughput, especially in challenged networks,
only anecdotal evidence has been provided showing the benefits for
time sensitive applications. 

\begin{figure}
\begin{centering}
\includegraphics[width=0.8\columnwidth]{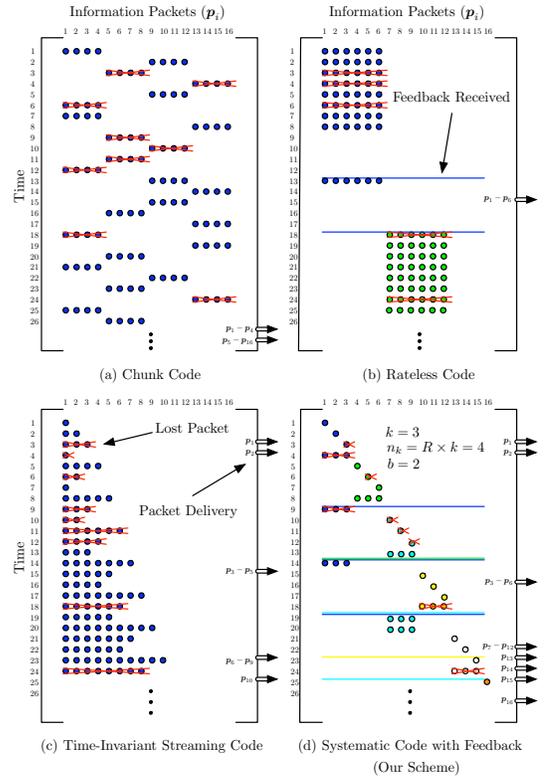}
\par\end{centering}

\caption{Coding matrices for various schemes assuming an identical loss pattern
and a feedback delay of 4 time-slots. The columns represent the original
information packet $\boldsymbol{p}_{i}$, and the rows represent the
composition of the transmitted packet at time $t$ (e.g., the transmitted
packet in time-slot 1 of (a) is $\boldsymbol{c}_{j}=\sum_{i=1}^{4}\alpha_{i,j}\boldsymbol{p}_{i}$
where $\alpha_{i,j}$ is defined in Section \ref{sec:Network-Model}).
Lines within a matrix indicate when feedback about a generation (represented
by different colors) was received.\label{fig:Coding-Examples}}
\vspace{-15pt}
\end{figure}

On the other hand, a large body of research investigating the delay
of coding in different settings has taken place. In general, most
of these works can be summarized by Figure \ref{fig:Coding-Examples}.
The coding delay of chunked and overlapping chunked codes (\cite{heidarzadeh_design_2012})
(shown in Figure \ref{fig:Coding-Examples}(a)), network coding in
time-division duplexing (TDD) channels (\cite{lucani_broadcasting_2009,lucani_online_2010,lucani_random_2009}),
and network coding in line networks where coding also occurs at intermediate
nodes (\cite{dikaliotis_delay_2009}) is well understood. In addition,
a non-asymptotic analysis of the delay distributions of random linear
network coding (RLNC) (\cite{nistor_non-asymptotic_2010}) and various
multicast scenarios (\cite{drinea_delay_2009,eryilmaz_delay_2006,swapna_throughput-delay_2013})
using a variant of the scheme in Figure \ref{fig:Coding-Examples}(b)
have also been investigated. Furthermore, the research that looks
at the in-order packet delay is provided in \cite{xia_analysis_2003}
and \cite{yao_multi-burst_2011} for uncoded systems, while \cite{nistor_network_2010},
\cite{sundararajan_feedback-based_2009}, and \cite{zeng_joint_2012}
considers the in-order packet delay for non-systematic coding schemes
similar to the one shown in Figure \ref{fig:Coding-Examples}(b).
However, these non-systematic schemes may not be the optimum strategy
in networks or communication channels with a long $RTT$.

Possibly the closest work to ours is that done by Joshi et. al. \cite{joshi_playback_2012,joshi_effect_2014}
and T\"{o}m\"{o}sk\"{o}zi et. al. \cite{toemoeskoezi_delay_2014}. Bounds on the
expected in-order delay and a study of the rate/delay trade-offs using
a time-invariant coding scheme is provided in \cite{joshi_playback_2012}
and \cite{joshi_effect_2014} where they assume feedback is instantaneous,
provided in a block-wise manner, or not available at all. A generalized
example of their coding scheme is shown in Figure \ref{fig:Coding-Examples}(c).
While their analysis provides insight into the benefits of coding
for streaming applications, their model is similar to a half-duplex
communication channel where the sender transmits a finite block a
information and then waits for the receiver to send feedback. Unfortunately,
it is unclear if their analysis can be extended to full-duplex channels
or models where feedback does not provide complete information about
the receiver's state-space. Finally, the work in \cite{toemoeskoezi_delay_2014}
considers the in-order delay of online network coding where feedback
determines the source packets used to generate coded packets. However,
they only provide experimental results and do not attempt an analysis.\vspace{-6pt}

\section{Coding Algorithm and System Model\label{sec:Network-Model}}

We consider a time-slotted model to determine the coding window size
$k$ and added redundancy $R\geq1$ that minimizes the per-packet,
or playback, delay $D$. The duration of each slot is $t_{s}=\nicefrac{s_{pkt}}{Rate}$
where $s_{pkt}$ is the size of each transmitted packet and $Rate$
is the transmission rate of the network. Also let $t_{p}$ be the
propagation delay between the sender and the receiver (i.e., $RTT=t_{s}+2t_{p}$
assuming that the size of each acknowledgement is sufficiently small).
Packet erasures are assumed to be independently and identically distributed
(i.i.d.) with $\epsilon$ being the probability of a packet erasure
within the network. 

Source packets $\boldsymbol{p}_{i}$, $i=\{1,\ldots,N\}$, are first
partitioned into coding generations $\boldsymbol{G}_{j}=\left\{ \boldsymbol{p}_{\left(j-1\right)k+1},\ldots,\boldsymbol{p}_{\min\left(jk,N\right)}\right\} $,
$j\in\left[1,\lceil\nicefrac{N}{k}\rceil\right]$. Each generation
$\boldsymbol{G}_{j}$ is transmitted using the systematic network
coding scheme shown in Algorithm \ref{alg:Code-Generation-Algorithm}
where the coding window spans the entire generation. The coded packets
$\boldsymbol{c}_{j,m}$ shown in the algorithm are generated by taking
random linear combinations of the $k$ information packets contained
within the generation where the coding coefficients $\alpha_{i,j,m}\in\mathbb{F}_{q}$
are chosen at random and $\boldsymbol{p}_{i}$ is treated as a vector
in $\mathbb{F}_{q}$. Once every packet in $\boldsymbol{G}_{j}$ has
been transmitted (both uncoded and coded), the coding window slides
to the next generation $\boldsymbol{G}_{j+1}$ and the process repeats
without waiting for feedback. 

\begin{algorithm}
\begin{algorithmic}[1]
\FORALL{$j\in\left[1,\lceil\nicefrac{N}{k}\rceil\right]$}

\forbody{\FORALL{Packets $\boldsymbol{p}_{i}$, $i\in\left[\left(j-1\right)k+1,\min\left(jk,N\right)\right]$}

\forbody{\STATE{Transmit $\boldsymbol{p}_{i}$}}

\FORALL{$m\in\left[1,n_{k}-k\right]$}

\forbody{\STATE{Transmit $\boldsymbol{c}_{j,m}=\sum_{i=\left(j-1\right)k+1}^{\min\left(jk,N\right)}\alpha_{i,j,m}\boldsymbol{p}_{i}$}}}
\end{algorithmic}

\caption{Code Generation Algorithm\label{alg:Code-Generation-Algorithm}}

\end{algorithm}

We assume that delayed feedback is provided about each generation
(i.e., multiple coding generations can be in-flight at any time);
and this feedback contains the number of degrees of freedom ($dofs$)
$l$ still required to decode the generation. If $l>0$, an additional
$n_{l}=R\times l\geq l$ coded packets (or $dofs$) are retransmitted.
This process is shown in Algorithm \ref{alg:-Retransmission-Algorithm}
and continues until all $k$ $dofs$ have been received and the generation
can be decoded and delivered.

\begin{algorithm}
\begin{algorithmic}[1]
\STATE{$ACK$ From $G_{j}$ Received}

\IF{No packets from $G_{j}$ in-flight \AND{} $l>0$}

\ifbody{\FORALL{$m\in\left[1,R\times l\right]$}

\forbody{\STATE{Transmit $\boldsymbol{c}_{j,m}=\sum_{i=\left(j-1\right)k+1}^{\min\left(jk,N\right)}\alpha_{i,j,m}\boldsymbol{p}_{i}$}}}
\end{algorithmic}

\caption{$DOF$ Retransmission Algorithm\label{alg:-Retransmission-Algorithm}}
\end{algorithm}

Figure \ref{fig:Coding-Examples}(d) provides an example of the proposed
scheme. Here we can see that source packets are partitioned into coding
generations of size $k=3$ packets, and one coded packet is also transmitted
for each generation (i.e., $R=1.33$). In this case, the first two
packets of the blue generation can be delivered, but the third packet
cannot since it is lost and the generation cannot be decoded. Delayed
feedback indicates that additional $dofs$ are needed and two additional
transmission attempts are required to successfully transmit the required
$dof$. Once the $dof$ is delivered, the remainder of the blue generation,
as well as the entire green generation, can be delivered in-order.

Due to the complexity of the process, several assumptions are needed.
First, retransmissions occur immediately after feedback is obtained
indicating additional $dofs$ are needed without waiting for the coding
window to shift to a new generation. Second, the time to transmit
packets after the first round does not increase the delay. For example,
the packet transmission time is $t_{s}$ seconds. Assuming $l$ $dof$
retransmissions are needed, the additional $n_{l}t_{s}$ seconds needed
to transmit these packets are not taken into account. Third, the number
of previously transmitted generations that can cause head-of-line
blocking is limited to $b-1$ where $b=\lceil\nicefrac{BDP}{n_{k}}\rceil$.
Fourth, all packets within a generation are available to the transport
layer without delay (i.e., we assume an infinite packet source). Finally,
the coding window/generation size with the added redundancy is smaller
than the $BDP$ (i.e., $n_{k}<BDP$). Without this assumption, feedback
will be received prior to the transmission of the coded packets allowing
for the use of SR-ARQ without a large impact to the performance. 

It is important to note that these assumptions provide a lower bound.
The first two assumptions ensure feedback is acted upon immediately
and does not impact the delay experienced by other generations. The
third assumption limits the possibility of a previously transmitted
generation preventing delivery, thereby decreasing the overall delay.

\section{Preliminaries\label{sec:Preliminaries}}

We first define several probability distributions and random variables
that will be used extensively in later sections. Define $\left[P\right]\in\mathbb{R}^{k+1\times k+1}$
to be the transition matrix of a Markov chain. Each transition within
the chain represents the number of $dofs$, or packets, successfully
received after a round of transmissions, and each state represents
the number of $dofs$ still needed by the client to decode. As a result,
the elements of $\left[P\right]$ can be defined as follows:\vspace{-1pt}
\begin{equation}
\left[P_{ij}\right]=\begin{cases}
B\left(n_{i},i-j,1-\epsilon\right) & \text{for }i\in[1,k],0<j\leq i\\
\sum_{m=i}^{n_{i}}B\left(n_{i},m,1-\epsilon\right) & \text{for }i\in[1,k],j=0\\
1 & \text{for }i=0,j=0,
\end{cases}\label{eq:a_ij-1}
\end{equation}
\vspace{-1pt}where $B(n,k,p)={n \choose k}p^{k}(1-p)^{n-k}$. Let
$X_{r}$ be the state of the chain at time $r$. It follows that $\text{Pr}\{X_{r}=j|X_{0}=i\}=\left[P_{ij}^{r}\right]$
for $r\geq1$ and $\left[P_{ij}^{0}\right]=0$. In our model, $X_{0}=k$
with probability equal to 1 and a generation is successfully decoded
when state $0$ is entered at time $r\geq1$. Furthermore, the probability
$\left[P_{i0}^{r}\right]$ is the probability that all packets within
a single generation have been successfully received in or before $r$
transmission rounds.

Using this Markov chain, define $Y$ to be the number of transmission
rounds required to transfer a single generation. The distribution
on $Y$ is:\vspace{-1pt} 
\begin{equation}
p_{Y}(y)=\begin{cases}
\left[P_{k0}^{y}\right]-\left[P_{k0}^{y-1}\right] & \text{for }y\geq1\\
0 & \text{otherwise.}
\end{cases}\label{eq:p_y}
\end{equation}
\vspace{-1pt}Next, define $Z_{i}$ to be the number of transmission
rounds required to transfer $i$ generations. Before defining the
distribution on $Z_{i}$, we first provide the following Lemma.
\begin{lem}
Let $N$ independent processes defined by the transition matrix $\left[P\right]$
start at the same time. The probability that all processes complete
in less than or equal to $z$ rounds, or transitions, with at least
one process completing in round $z$ is $\text{Pr}\{Z=z\}=\left[P_{k0}^{z}\right]^{N}-\left[P_{k0}^{z-1}\right]^{N}$.\label{lem:ProbNBlocks}\end{lem}
\begin{IEEEproof}
Let $f\left(z\right)=1-\sum_{j=1}^{z-1}p_{Y}\left(j\right)$. The
probability of $N$ independent processes completing in less than
or equal to $z$ rounds with at least one process completing in round
$z$ is:\vspace{-1pt}

\begin{alignat}{1}
\text{Pr}\{Z=z\} & =\sum_{i=1}^{N}B\left(N,i,f\left(z\right)\right)\left(\frac{p_{Y}\left(z\right)}{f\left(z\right)}\right)^{i}\\
 & =\sum_{i=1}^{N}{N \choose i}\left(p_{Y}\left(z\right)\right)^{i}\left[P_{k0}^{z-1}\right]^{N-i}\\
 & =\frac{\left[P_{k0}^{z-1}\right]^{N+1}\left(\left[P_{k0}^{z}\right]^{N+1}-\left[P_{k0}^{z-1}\right]^{N+1}\right)}{\left[P_{k0}^{z-1}\right]^{N+1}\left[P_{k0}^{z}\right]}\nonumber \\
 & -\frac{\left[P_{k0}^{z-1}\right]^{2N+1}\left(\left[P_{k0}^{z}\right]-\left[P_{k0}^{z-1}\right]\right)}{\left[P_{k0}^{z-1}\right]^{N+1}\left[P_{k0}^{z}\right]}\\
 & =\left[P_{k0}^{z}\right]^{N}-\left[P_{k0}^{z-1}\right]^{N}.
\end{alignat}
\vspace{-1pt}
\end{IEEEproof}
Given Lemma \ref{lem:ProbNBlocks}, the distribution on $Z_{i}$ is:
\begin{equation}
p_{Z_{i}}(z_{i})=\begin{cases}
\left[P_{k0}^{z_{i}}\right]^{i}-\left[P_{k0}^{z_{i}-1}\right]^{i} & \text{for }z_{i}\geq1,i\leq b-1\\
0 & \text{otherwise.}
\end{cases}\label{eq:p_z}
\end{equation}

Also define $S$ to be the number of uncoded packets that are successfully
transferred within a generation prior to the first packet loss. The
distribution on $S$ is:
\begin{equation}
p_{S|Y}(s|y)=\begin{cases}
\epsilon\left(1-\epsilon\right)^{s} & \text{for }s\in[0,k-1],y=1\\
\left(1-\epsilon\right)^{s} & \text{for }s=k,y=1\\
\frac{\epsilon\left(1-\epsilon\right)^{s}}{1-\left(1-\epsilon\right)^{k}} & \text{for }s\in[0,k-1],y\neq1,\\
0 & \text{otherwise},
\end{cases}\label{eq:p_s-1}
\end{equation}
and its first three moments are given by the following lemma.
\begin{lem}
\label{lem:exp_s}Define $\overline{s_{1}^{i}}=\mathbb{E}\left[S^{i}|Y=1\right]$
and $\overline{s_{2}^{i}}=\mathbb{E}\left[S^{i}|Y\neq1\right]$. Then
given $Y=y$, the first through third moments of $S$ are
\begin{alignat}{1}
\overline{s_{1}^{1}} & =\frac{1-\epsilon}{\epsilon}\left(1-\left(1-\epsilon\right)^{k}\right),\label{eq:exp_s-1}\\
\overline{s_{1}^{2}} & =\frac{2\left(1-\epsilon\right)}{\epsilon^{2}}\left(1-\left(k\epsilon+1\right)\left(1-\epsilon\right)^{k}\right)-\overline{s_{1}^{1}},\label{eq:exp_s2-1}\\
\overline{s_{1}^{3}} & =\frac{6\left(1-\epsilon\right)^{3}}{\epsilon^{3}}\left(1-\left(k\epsilon+1\right)(1-\epsilon)^{k}\right)+3\left(1-\epsilon\right)\overline{s_{1}^{2}}\nonumber \\
 & -\frac{3k}{\epsilon}\left(k+1\right)\left(1-\epsilon\right)^{k+1}+\left(4-3\epsilon\right)\overline{s_{1}^{1}},\label{eq:exp_s3-1}
\end{alignat}

and
\begin{equation}
\overline{s_{2}^{i}}=\frac{\overline{s_{1}^{i}}-k^{i}\left(1-\epsilon\right)^{k}}{1-\left(1-\epsilon\right)^{k}},\label{eq:exp_s-2}
\end{equation}
for $i=1,2,3$.\end{lem}
\begin{IEEEproof}
Define the moment generating function of $S$ when $Y=1$ to be 
\begin{eqnarray}
M_{S|Y}(t) & = & \mathbb{E}[e^{tS}|Y=1]\\
 & = & \frac{\epsilon\left(1-e^{kt}\left(1-\epsilon\right)^{k}\right)}{1-e^{t}+\epsilon e^{t}}+e^{kt}\left(1-\epsilon\right)^{k}.\label{eq:moment-gen-s}
\end{eqnarray}
The first, second, and third moments of $S$ when $Y=1$ are then
$\nicefrac{\delta}{\delta t}M_{S|Y}(0)$, $\nicefrac{\delta^{2}}{\delta t^{2}}M_{S|Y}(0)$,
and $\nicefrac{\delta^{3}}{\delta t^{3}}M_{S|Y}(0)$ respectively. 

For $Y\neq1$, we need to scale the above expectations accordingly.
This can be done by subtracting the term $k^{i}\left(1-\epsilon\right)^{k}$
from each of the moments above and dividing by $1-\left(1-\epsilon\right)^{k}$.
\end{IEEEproof}
Finally, let $V_{i}$, $i\leq b-1$, describe the position of the
last received generation preventing delivery in round $z_{i}$. The
following lemma helps to define the distribution on $V_{i}$.
\begin{lem}
\label{lem:exp_v}Let $N$ independent processes defined by the transition
matrix $\left[P\right]$ start at the same time, and all processes
complete in or before round $z_{N}$ with at least one process completing
in round $z_{N}$. The probability that the $j$th process is the
last to complete is defined by the distribution
\begin{equation}
p_{V_{N}|Z_{N}}\left(v_{N}|z_{N}\right)=\frac{\left[P_{k0}^{z_{N}}\right]^{N-v_{N}-1}\left[P_{k0}^{z_{N}-1}\right]^{v_{N}}p_{Y}\left(z_{N}\right)}{p_{Z_{N}}(z_{N})}
\end{equation}
for $v_{N}=0,\ldots N-1$, $j=N-v_{N}$, $p_{Y}(y)$ defined in (\ref{eq:p_y}),
and $p_{Z_{i}}\left(z_{i}\right)$ defined in (\ref{eq:p_z}). Furthermore,
define $\overline{v_{N}^{i}}=\mathbb{E}\left[V_{N}^{i}|Z_{N}\right]$.
Then 
\begin{equation}
\overline{v_{N}^{1}}=\frac{\left[P_{k0}^{z_{N}-1}\right]}{p_{Y}\left(z_{N}\right)}-\frac{N\left[P_{k0}^{z_{N}-1}\right]^{N}}{p_{Z_{N}}\left(z_{N}\right)},\label{eq:exp_v}
\end{equation}
and
\begin{eqnarray}
\overline{v_{N}^{2}} & = & \frac{\left[P_{k0}^{z_{N}-1}\right]}{p_{Y}\left(z_{N}\right)}+\frac{2\left[P_{k0}^{z_{N}-1}\right]^{2}}{p_{Y}^{2}\left(z_{N}\right)}-\frac{N^{2}\left[P_{k0}^{z_{N}-1}\right]^{N}}{p_{Z_{N}}\left(z_{N}\right)}\nonumber \\
 &  & -\frac{2N\left[P_{k0}^{z_{N}-1}\right]^{N+1}}{p_{Y}\left(z_{N}\right)p_{Z_{N}}\left(z_{N}\right)}.
\end{eqnarray}
\end{lem}
\begin{IEEEproof}
Let $\beta_{z_{N}}=\nicefrac{p_{Y}\left(z_{N}\right)}{\left[P_{k0}^{z_{N}}\right]}$,
be the probability of a generation finishing in round $z_{N}$ given
all of the $N$ generations have completed transmission in or before
round $z_{N}$. The distribution on $V_{N}\in[0,N-1]$ is 
\begin{alignat}{1}
p_{V_{N}|Z_{N}}\left(v_{N}|z_{N}\right) & =\frac{\beta_{z_{N}}\left(1-\beta_{z_{N}}\right)^{v_{N}}}{\sum_{j=0}^{N-1}\beta_{z_{N}}\left(1-\beta_{z_{N}}\right)^{j}}\\
 & =\frac{\beta_{z_{N}}\left(1-\beta_{z_{N}}\right)^{v_{N}+1}}{1-\left(1-\beta_{z_{N}}\right)^{N}}\\
 & =\frac{\left[P_{k0}^{z_{N}}\right]^{N-v_{N}-1}\left[P_{k0}^{z_{N}-1}\right]^{v_{N}}p_{Y}\left(z_{N}\right)}{p_{Z_{N}}(z_{N})}.
\end{alignat}
Define the moment generating function of $V_{N}$ given $Z_{N}$ to
be 
\begin{alignat}{1}
M_{V_{N}|Z_{N}}(t) & =\mathbb{E}[e^{tV_{N}}|Z_{N}=z_{N}]\nonumber \\
 & =\frac{\left(\left[P_{k0}^{z_{N}}\right]^{N}-\left[P_{k0}^{z_{N}-1}\right]^{N}e^{Nt}\right)p_{Y}(z_{N})}{\left(\left[P_{k0}^{z_{N}}\right]-\left[P_{k0}^{z_{N}-1}\right]e^{t}\right)p_{Z_{N}}(z_{N})}.
\end{alignat}
The first and second moments of $V_{N}$ given $Z_{N}$ are $\nicefrac{\delta}{\delta t}M_{V_{N}|Z_{N}}(0)$
and $\nicefrac{\delta^{2}}{\delta t^{2}}M_{V_{N}|Z_{N}}(0)$ respectively.
\end{IEEEproof}
Now that we have the distributions for the random variables $Y$,
$Z_{i}$, $S$, and $V_{i}$, as well as several relevant moments,
we have the tools needed to derive the expected in-order delivery
delay.

\section{Expected In-Order Delivery Delay\label{sec:Expected-In-Order-Delivery}}

A lower bound on the expected delay, $\mathbb{E}[D]$, can be derived
using the law of total expectation: 
\begin{eqnarray}
\mathbb{E}[D] & = & \mathbb{E}_{Y}\left[\mathbb{E}_{Z_{b-1}}\left[\mathbb{E}_{D}\left[D|Y,Z_{b-1}\right]\right]\right].\label{eq:exp_d}
\end{eqnarray}
From (\ref{eq:exp_d}), there are four distinct cases that must be
evaluated. For each case, define $\bar{d}_{Y=y,Z=z}=\mathbb{E}\left[D|Y=y,Z_{b-1}=z\right]$.

\subsection{Case 1: $Y=1,Z_{b-1}=1$ }

\begin{figure}
\begin{centering}
\includegraphics[width=0.85\columnwidth]{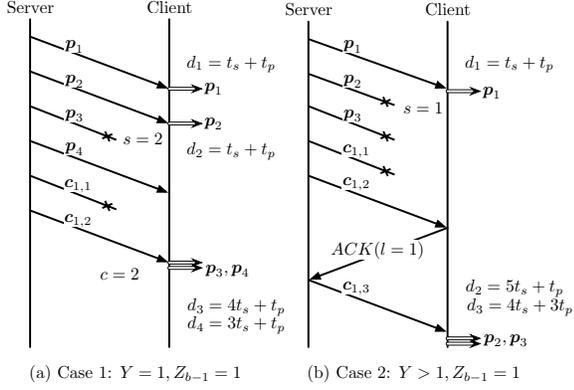}
\par\end{centering}

\caption{Example of (a) case 1 and (b) case 2. The delay $d_{i}$ of each packet
is listed next to the time when it is delivered to the application
layer.\label{fig:Example-Cases-1-and-2}}
\vspace{-15pt}
\end{figure}

The latest generation in transit completes within the first round
of transmission and no previously transmitted generations prevent
delivery. As a result, all packets received prior to the first loss
(i.e., packets $\boldsymbol{p}_{1},\ldots\boldsymbol{p}_{s}$) are
immediately delivered. Once a packet loss is observed, packets received
after the loss (i.e., packets $\boldsymbol{p}_{s+1},\ldots,\boldsymbol{p}_{k}$)
are buffered until the entire generation is decoded. An example is
given in Figure \ref{fig:Example-Cases-1-and-2}(a) where $n_{k}=6$,
$k=4$, the number of packets received prior to the first loss is
$s=2$, and the number of coded packets needed to recover from the
two packet losses is $c=2$. Taking the expectation over all $S$
and all packets within the generation, the mean delay is\vspace{-10pt}

\begin{alignat}{1}
\bar{d}_{Y=1,Z=1} & =\sum_{s=0}^{k-1}\left(\left(t_{s}+t_{p}\right)\frac{s}{k}+\frac{1}{k}\sum_{i=0}^{k-s-1}\biggl(t_{p}+\bigl(k-s-i\right.\nonumber \\
 & +\mathbb{E}\left[C|S\right]\bigr)t_{s}\biggr)\Biggr)p_{S|Y}\left(s|1\right)+\left(t_{s}+t_{p}\right)p_{S|Y}\left(k|1\right)\\
 & =t_{p}+\frac{t_{s}}{2k}\biggl(\overline{s_{1}^{2}}-\left(2k-1\right)\overline{s_{1}^{1}}+k\left(k+1\right)\nonumber \\
 & +2\sum_{s=0}^{k-1}\left(k-s\right)\mathbb{E}\left[C|S\right]p_{S|Y}\left(s|1\right)\biggr)\\
 & \geq t_{p}+\frac{t_{s}}{2k}\biggl(\overline{s_{1}^{2}}-\left(2k-1\right)\overline{s_{1}^{1}}+k\left(k+1\right)\nonumber \\
 & +2\sum_{s=0}^{k-1}\left(k-s\right)p_{S|Y}\left(s|1\right)\biggr)\label{eq:case-1-bound}\\
 & =\frac{t_{s}}{2k}\biggl(\overline{s_{1}^{2}}-\left(2k+1\right)\overline{s_{1}^{1}}+k\left(k+3\right)\biggr)+t_{p},\label{eq:case-1}
\end{alignat}
where $\overline{s_{1}^{1}}$ and $\overline{s_{1}^{2}}$ are given
by Lemma \ref{lem:exp_s}; and $\mathbb{E}\left[C|S\right]$ is the
expected number of coded packets needed to recover from all packet
erasures occurring in the first $k$ packets. When $s<k$, the number
of coded packets required is at least one (i.e., $\mathbb{E}\left[C|S\right]\geq1$)
leading to the bound in (\ref{eq:case-1-bound}).

\subsection{Case 2: $Y>1,Z_{b-1}=1$}

All packets $\left\{ \boldsymbol{p}_{1},\ldots\boldsymbol{p}_{s}\right\} $
are delivered immediately until the first packet loss is observed.
Since $Y>1$, at least one retransmission event is needed to properly
decode. Once all $k$ $dofs$ have been received and the generation
can be decoded, the remaining packets $\left\{ \boldsymbol{p}_{s+1},\ldots,\boldsymbol{p}_{k}\right\} $
are delivered in-order. An example is provided in Figure \ref{fig:Example-Cases-1-and-2}(b).
The generation cannot be decoded because there are too many packet
losses during the first transmission attempt. As a result, one additional
$dof$ is retransmitted allowing the client to decode in round two
(i.e., $Y=2$). Taking the expectation over all $S$ and all packets
within the generation, the expected delay for this case is\vspace{-15pt}

\begin{alignat}{1}
\bar{d}_{Y>1,Z=1} & =\frac{1}{k}\sum_{s=0}^{k-1}\biggl(\left(t_{p}+t_{s}\right)s+\sum_{i=0}^{k-s-1}\Bigl(t_{p}+2\left(y-1\right)t_{p}\nonumber \\
 & +\left(k-s-i+n_{k}-k\right)t_{s}\Bigr)\biggr)p_{S|Y}(s|y\neq1)\\
 & =\biggl(\frac{1}{2k}\overline{s_{2}^{2}}-\frac{1}{2k}\left(2n_{k}-1\right)\overline{s_{2}^{1}}+n_{k}-\frac{1}{2}k+\frac{1}{2}\biggr)t_{s}\nonumber \\
 & -\biggl(\frac{2}{k}\left(y-1\right)\overline{s_{2}^{1}}-2y+1\biggr)t_{p},\label{eq:case-2}
\end{alignat}
where $\overline{s_{2}^{1}}$ and $\overline{s_{2}^{2}}$ are given
by Lemma \ref{lem:exp_s}. It is important to note that we do not
take into account the time to transmit packets after the first round
(see the assumptions in Section \ref{sec:Network-Model}).

\subsection{Case 3: $Z_{b-1}>Y\geq1,Z_{b-1}>1$}

\begin{figure}
\begin{centering}
\includegraphics[width=0.9\columnwidth]{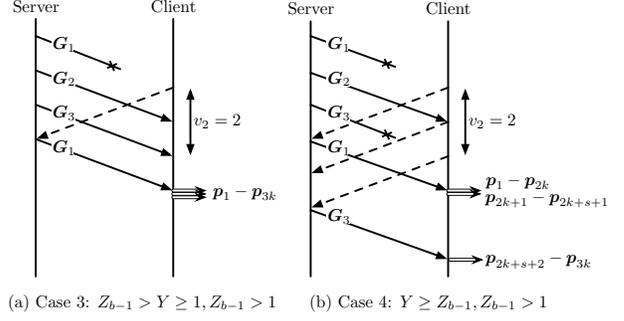}
\par\end{centering}

\caption{Example of (a) case 3 and (b) case 4 where $b=3$.\label{fig:Example-Case3-Case4}}
\vspace{-20pt}
\end{figure}

In this case, generation $\boldsymbol{G}_{j}$ completes prior to
a previously sent generation. As a result, all packets $\left\{ \boldsymbol{p}_{\left(j-1\right)k+1},\dots,\boldsymbol{p}_{jk}\right\} \in\boldsymbol{G}_{j}$
are buffered until all previous generations have been delivered. Once
there are no earlier generations preventing in-order delivery, all
packets in $\boldsymbol{G}_{j}$ are immediately delivered. Figure
\ref{fig:Example-Case3-Case4}(a) provides an example. Consider the
delay experienced by packets in $\boldsymbol{G}_{3}$. While $\boldsymbol{G}_{3}$
is successfully decoded after the first transmission attempt, generation
$\boldsymbol{G}_{1}$ cannot be decoded forcing all packets in $\boldsymbol{G}_{3}$
to be buffered until $\boldsymbol{G}_{1}$ is delivered. Taking the
expectation over all packets within the generation and all possible
locations of the last unsuccessfully decoded generation, the expected
delay is\vspace{-5pt}
\begin{alignat}{1}
\bar{d}_{Z>Y\geq1,Z>1} & =\frac{1}{k}\sum_{i=1}^{k}\biggl(\left(n_{k}-k+i\right)t_{s}+t_{p}+2t_{p}\left(z-1\right)\nonumber \\
 & -\left(\overline{v_{b-1}^{1}}+1\right)n_{k}t_{s}\biggr)\\
 & =\left(2z-1\right)t_{p}-\left(\overline{v_{b-1}^{1}}n_{k}+\frac{k-1}{2}\right)t_{s},\label{eq:case-4}
\end{alignat}
where $\overline{v_{b-1}^{1}}$ is given by Lemma \ref{lem:exp_v}.

\subsection{Case 4: $Y\geq Z_{b-1},Z_{b-1}>1$}

Finally, this case is a mixture of the last two. The generation $\boldsymbol{G}_{j}$
completes after all previously transmitted generations, but it requires
more than one transmission round to decode. Packets received before
the first packet loss are buffered until all previous generations
are delivered, and packets received after the first packet loss are
buffered until $\boldsymbol{G}_{j}$ can be decoded. An example is
provided in Figure \ref{fig:Example-Case3-Case4}(b). Consider the
delay of packets in $\boldsymbol{G}_{3}$. Both $\boldsymbol{G}_{1}$
and $\boldsymbol{G}_{3}$ cannot be decoded after the first transmission
attempt. After the second transmission attempt, $\boldsymbol{G}_{1}$
can be decoded allowing packets $\left\{ \boldsymbol{p}_{2k+1},\ldots,\boldsymbol{p}_{2k+s+1}\right\} \in\boldsymbol{G}_{3}$
to be delivered; although packets $\left\{ \boldsymbol{p}_{2k+s+2},\ldots,\boldsymbol{p}_{3k}\right\} \in\boldsymbol{G}_{3}$
must wait to be delivered until after $\boldsymbol{G}_{3}$ is decoded.
Taking the expectation over all $S$, all packets within the generation,
and all possible locations of the last unsuccessfully decoded generation,
the expected delay is\vspace{-2pt}
\begin{alignat}{1}
\bar{d}_{Y\geq Z,Z>1} & =\frac{1}{k}\sum_{s=0}^{k-1}\Biggl(\sum_{i=1}^{s}\biggl(\left(n_{k}-i+1\right)t_{s}+2t_{p}\left(z-\frac{1}{2}\right)\nonumber \\
 & -\left(\overline{v_{b-1}^{1}}+1\right)n_{k}t_{s}\biggr)+\sum_{j=s+1}^{k}\Bigl(2t_{p}\Bigl(y-\frac{1}{2}\Bigr)\nonumber \\
 & +\left(n_{k}-j+1\right)t_{s}\Bigr)\Biggr)p_{S|Y}(s|y\neq1)\\
 & =\biggl(\frac{2\left(z-y\right)}{k}\overline{s_{2}^{1}}+2y-1\biggr)t_{p}\nonumber \\
 & -\biggl(\frac{n_{k}}{k}\left(\overline{v_{b-1}^{1}}+1\right)\overline{s_{2}^{1}}-n_{k}+\frac{1}{2}k-\frac{1}{2}\biggr)t_{s}.\label{eq:case-3}
\end{alignat}
The expectations $\overline{s_{2}^{1}}$ and $\overline{v_{b-1}^{1}}$
are given by Lemmas \ref{lem:exp_s} and \ref{eq:exp_v} respectively.

Combining the cases above, we obtain the following:\vspace{-2pt}
\begin{thm}
The expected in-order packet delay for the proposed coding scheme
is lower bounded by 
\begin{equation}
\mathbb{E}[D]\geq\sum_{z_{b-1}\geq1}\sum_{y\geq1}\bar{d}_{Y=y,Z=z_{b-1}}p_{Y}(y)p_{Z_{b-1}}(z_{b-1}).\label{eq:exp_d-2}
\end{equation}
\vspace{-2pt}where $\bar{d}_{Y=y,Z=z_{b-1}}$ is given in equations
(\ref{eq:case-1}), (\ref{eq:case-2}), (\ref{eq:case-4}), and (\ref{eq:case-3});
and the distributions $p_{Y}(y)$ and $p_{Z_{b-1}}(z_{b-1})$ are
given in equations (\ref{eq:p_y}) and (\ref{eq:p_z}) respectively.\vspace{-1pt}
\end{thm}

\section{In-Order Delivery Delay Variance\label{sec:In-Order-Delivery-Delay}}

The second moment of the in-order delivery delay can be determined
in a similar manner as the first. Again, we can use the law of total
expectation to find the moment:\vspace{-5pt}

\begin{eqnarray}
\mathbb{E}[D^{2}] & = & \mathbb{E}_{Y}\left[\mathbb{E}_{Z_{b-1}}\left[\mathbb{E}_{D}\left[D^{2}|Y,Z_{b-1}\right]\right]\right].\label{eq:exp_d2}
\end{eqnarray}
As with the first moment, four distinct cases exist that must be dealt
with separately. For each case, define $\bar{d}_{Y=y,Z=z}^{2}=\mathbb{E}\left[D^{2}|Y=y,Z_{b-1}=z\right]$.
While we omit the initial step in the derivation of each case, $\bar{d}_{Y,Z}^{2}$
can be determined using the same assumptions as above.\vspace{-1pt}

\subsection{Case 1: $Y=1,Z_{b-1}=1$}

Using the expectations defined in Lemma \ref{lem:exp_s}, the second
moment $\bar{d}_{Y=1,Z=1}^{2}$ is shown below. For $s<k$, the number
of coded packets needed to decode the generation will always be greater
than or equal to one (i.e., $c\geq1$). Therefore, the bound in (\ref{eq:case-1-2})
follows from letting $c=1$ and $p_{C|S}\left(1|s\right)=1$ for all
$s$. 

\begin{alignat}{1}
\bar{d}_{Y=1,Z=1}^{2} & \geq t_{p}^{2}+\left(k+3\right)t_{p}t_{s}+\frac{1}{6}\left(2k^{2}+9k+13\right)t_{s}^{2}\nonumber \\
 & -\biggl(\left(k+3+\frac{7}{6k}\right)t_{s}^{2}+\left(\frac{2k+1}{k}\right)t_{p}t_{s}\biggr)\overline{s_{1}^{1}}\nonumber \\
 & +\Biggl(\left(\frac{2k+3}{2k}\right)t_{s}^{2}+\frac{1}{k}t_{p}t_{s}\Biggr)\overline{s_{1}^{2}}-\frac{1}{3k}t_{s}^{2}\overline{s_{1}^{3}}.\label{eq:case-1-2}
\end{alignat}

\subsection{Case 2: $Y>1,Z_{b-1}=1$}

This case can be derived in a similar manner as the last. Again, each
$\overline{s_{2}^{i}}$, $i=\left\{ 1,2,3\right\} $, are given by
Lemma \ref{lem:exp_s}.

\begin{alignat}{1}
\bar{d}_{Y>1,Z=1}^{2}\nonumber \\
= & \biggl(n_{k}\left(n_{k}-k+1\right)+\frac{1}{6k}\left(2k^{3}-3k^{2}+k+6\right)\biggr)t_{s}^{2}\nonumber \\
+ & \biggl(2n_{k}\left(2y-1\right)-2y\left(k-1\right)+k-1+\frac{2}{k}\biggr)t_{p}t_{s}\nonumber \\
+ & \frac{1}{2k}\left(\left(2n_{k}+1\right)t_{s}^{2}+2\left(2y-1\right)t_{p}t_{s}\right)\overline{s_{2}^{2}}-\frac{1}{3k}t_{s}^{2}\overline{s_{2}^{3}}\nonumber \\
- & \frac{1}{k}\biggl((n_{k}^{2}+n_{k}+\frac{1}{6})t_{s}^{2}+\left(2y-1\right)\left(2n_{k}+1\right)t_{p}t_{s}\nonumber \\
+ & (2y-1)^{2}t_{p}^{2}\biggr)\overline{s_{2}^{1}}+\left(\left(2y-1\right)^{2}+\frac{1}{k}\right)t_{p}^{2}.\label{eq:case-2-2}
\end{alignat}

\subsection{Case 3: $Z_{b-1}>Y\geq1,Z_{b-1}>1$}

The second moment $\bar{d}_{Z>Y\geq1,Z>1}^{2}$ can be derived as
follows:

\begin{alignat}{1}
\bar{d}_{Z>Y\geq1,Z>1}^{2}\nonumber \\
= & \biggl(n_{k}^{2}\overline{v_{b-1}^{2}}+\left(k-1\right)\left(n_{k}\overline{v_{b-1}^{1}}+\frac{1}{3}k-\frac{1}{6}\right)\biggr)t_{s}^{2}\nonumber \\
- & (2z-1)\biggl(2n_{k}\overline{v_{b-1}^{1}}+k-1\biggr)t_{p}t_{s}+(2z-1)^{2}t_{p}^{2}.\label{eq:case-4-2}
\end{alignat}

\subsection{Case 4: $Y\geq Z_{b-1},Z_{b-1}>1$}

Using Lemmas \ref{lem:exp_s} and \ref{lem:exp_v}, the second moment
$\bar{d}_{Y\geq Z,Z>1}^{2}$ is:

\begin{alignat}{1}
\bar{d}_{Y\geq Z,Z>1}^{2}\nonumber \\
= & \left(2y-1\right)\biggl(\left(2n_{k}-k+1\right)t_{s}t_{p}+\left(2y-1\right)t_{p}^{2}\biggr)\nonumber \\
+ & \biggl(n_{k}\left(n_{k}-k+1\right)+\frac{1}{6}\left(2k^{2}-3k+1\right)\biggr)t_{s}^{2}\nonumber \\
+ & \frac{1}{k}\biggl(n_{k}\left(\overline{v_{b-1}^{1}}+1\right)t_{s}^{2}+2\left(y-z\right)t_{p}t_{s}\biggr)\overline{s_{2}^{2}}\nonumber \\
+ & \frac{1}{k}\biggl(n_{k}\Bigl(n_{k}\left(\overline{v_{b-1}^{2}}-1\right)-\overline{v_{b-1}^{1}}-1\Bigr)t_{s}^{2}\nonumber \\
- & 2\biggl(n_{k}\left(\overline{v_{b-1}^{1}}\left(2z-1\right)+\left(2y-1\right)\right)+y-z\biggr)t_{p}t_{s}\nonumber \\
- & 4\left(y-z\right)\left(y+z-1\right)t_{p}^{2}\biggr)\overline{s_{2}^{1}}.\label{eq:case-3-2}
\end{alignat}

Combining the cases above, we obtain:
\begin{thm}
The second moment of the in-order packet delay for the proposed coding
scheme is lower bounded by\vspace{-3pt}
\begin{equation}
\mathbb{E}[D^{2}]\geq\sum_{z_{b-1}\geq1}\sum_{y\geq1}\bar{d}_{Y=y,Z=z_{b-1}}^{2}p_{Y}(y)p_{Z_{b-1}}(z_{b-1}),\label{eq:exp_d2-2}
\end{equation}
where $\bar{d}_{Y=y,Z=z_{b-1}}^{2}$ is given in equations (\ref{eq:case-1-2}),
(\ref{eq:case-2-2}), (\ref{eq:case-4-2}), and (\ref{eq:case-3-2});
and the distributions $p_{Y}\left(y\right)$ and $p_{Z_{b-1}}\left(z_{b-1}\right)$
are given in (\ref{eq:p_y}) and (\ref{eq:p_z}) respectively. Furthermore,
the in-order delay variance is $\sigma_{D}^{2}=\mathbb{E}[D^{2}]-\mathbb{E}[D]^{2}$
where $\mathbb{E}\left[D\right]$ is given in (\ref{eq:exp_d}).\vspace{-8pt}
\end{thm}

\section{Efficiency\vspace{-2pt}\label{sec:Efficiency}}

The above results show adding redundancy into a packet stream decreases
the in-order delivery delay. However, doing so comes with a cost.
We characterize this cost in terms of efficiency. Before defining
the efficiency, let $M_{i}$, $i\in[0,k]$, be the number of packets
received at the sink as a result of transmitting a generation of size
$i$. Alternatively, $M_{i}$ is the total number of packets received
by the sink for any path starting in state $i$ and ending in state
$0$ of the Markov chain defined in Section \ref{sec:Network-Model}.
Furthermore, define $M_{ij}$ to be the number of packets received
by the sink as a result of a single transition from state $i$ to
state $j$ (i.e., $i\rightarrow j$). $M_{ij}$ is deterministic (e.g.,
$m_{ij}=i-j$) when $i,j\geq1$ and $i\geq j$. For any transition
$i\rightarrow0$, $i\geq1$, $m_{i0}\in[i,n_{i}]$ has probability\vspace{-3pt}
\begin{eqnarray}
p_{M_{i0}}(m_{i0}) & = & \frac{B\left(n_{i},m_{i0},1-\epsilon\right)}{\sum_{j=i}^{n_{i}}B\left(n_{i},j,1-\epsilon\right)}\\
 & = & \nicefrac{1}{a_{i0}}B\left(n_{i},m_{i0},1-\epsilon\right).
\end{eqnarray}
\vspace{-3pt}Therefore, the expected number of packets received by
the sink is\vspace{-3pt}
\begin{equation}
\mathbb{E}[M_{ij}]=\begin{cases}
i-j & \mbox{for }i,j\geq1,i\geq j\\
\frac{1}{a_{i0}}\sum_{x=i}^{n_{i}}x\cdot B\left(n_{i},x,1-\epsilon\right) & \mbox{for }i\geq1,j=0.
\end{cases}
\end{equation}
\vspace{-3pt}

Given $\mathbb{E}[M_{ij}]$ $\forall i,j$, the total number of packets
received by the sink when transmitting a generation of size $i$ is
\begin{equation}
\mathbb{E}\left[M_{i}\right]=\frac{1}{1-a_{ii}}\left(\sum_{j=0}^{i-1}\left(\mathbb{E}\left[M_{ij}\right]+\mathbb{E}\left[M_{j}\right]\right)a_{ij}\right)
\end{equation}
\vspace{-3pt}where $\mathbb{E}[M_{0}]=0$. This leads us to the following
theorem.
\begin{thm}
The efficiency $\eta_{k}$, defined as the ratio between the number
of information packets or $dofs$ within each generation of size $k$
and the expected number of packets received by the sink, is\vspace{-3pt}
\begin{equation}
\eta_{k}\triangleq\frac{k}{\mathbb{E}\left[M_{k}\right]}.\label{eq:efficiency}
\end{equation}

\end{thm}

\section{Numerical Results\label{sec:Numerical-Results}}

The analysis presented in the last few sections provided a method
to lower bound the expected in-order delivery delay. Unfortunately,
the complexity of the process prevents us from determining a closed
form expression for this bound. However, this section will provide
numerical results. Before proceeding, several items need to be noted.
First, we do not consider the terms where $p_{Y}\left(y\right)p_{Z_{b-1}}\left(z_{b-1}\right)<10^{-6}$
when calculating $\mathbb{E}[D]$ and $\mathbb{E}\left[D^{2}\right]$
since they have little effect on the overall calculation. Second,
the analytical curves are sampled at local maxima. As the code generation
size increases, the number of in-flight generations, $b=\lceil\nicefrac{BDP}{k}\rceil$,
incrementally decreases. Upon each decrease in $b$, a discontinuity
occurs that causes an artificial decrease in $\mathbb{E}[D]$ that
becomes less noticeable as $k$ increases towards the next decrease
in $b$. This transient behavior in the analysis is more prominent
in the cases where $R\approx\nicefrac{1}{1-\epsilon}$ and less so
when $R\gg\nicefrac{1}{1-\epsilon}$. Regardless, the figures show
an approximation with this transient behavior removed. Third, we note
that $Rk$ may not be an integer. To overcome this issue when generating
and transmitting coded packets, $\lceil Rk\rceil-k$ and $\lfloor Rk\rfloor-k$
coded packets are sent with probability $Rk-\lfloor Rk\rfloor$ and
$\lceil Rk\rceil-Rk$ respectively $ $Finally, we denote the redundancy
used in each of the figures as $R_{x}=\nicefrac{\left(1+x\right)}{\left(1-\epsilon\right)}$.

\subsection{Coding Window Size and Redundancy Selection}

\begin{figure}
\begin{centering}
\subfloat[$\epsilon=0.01$]{\begin{centering}
\includegraphics[width=1\columnwidth]{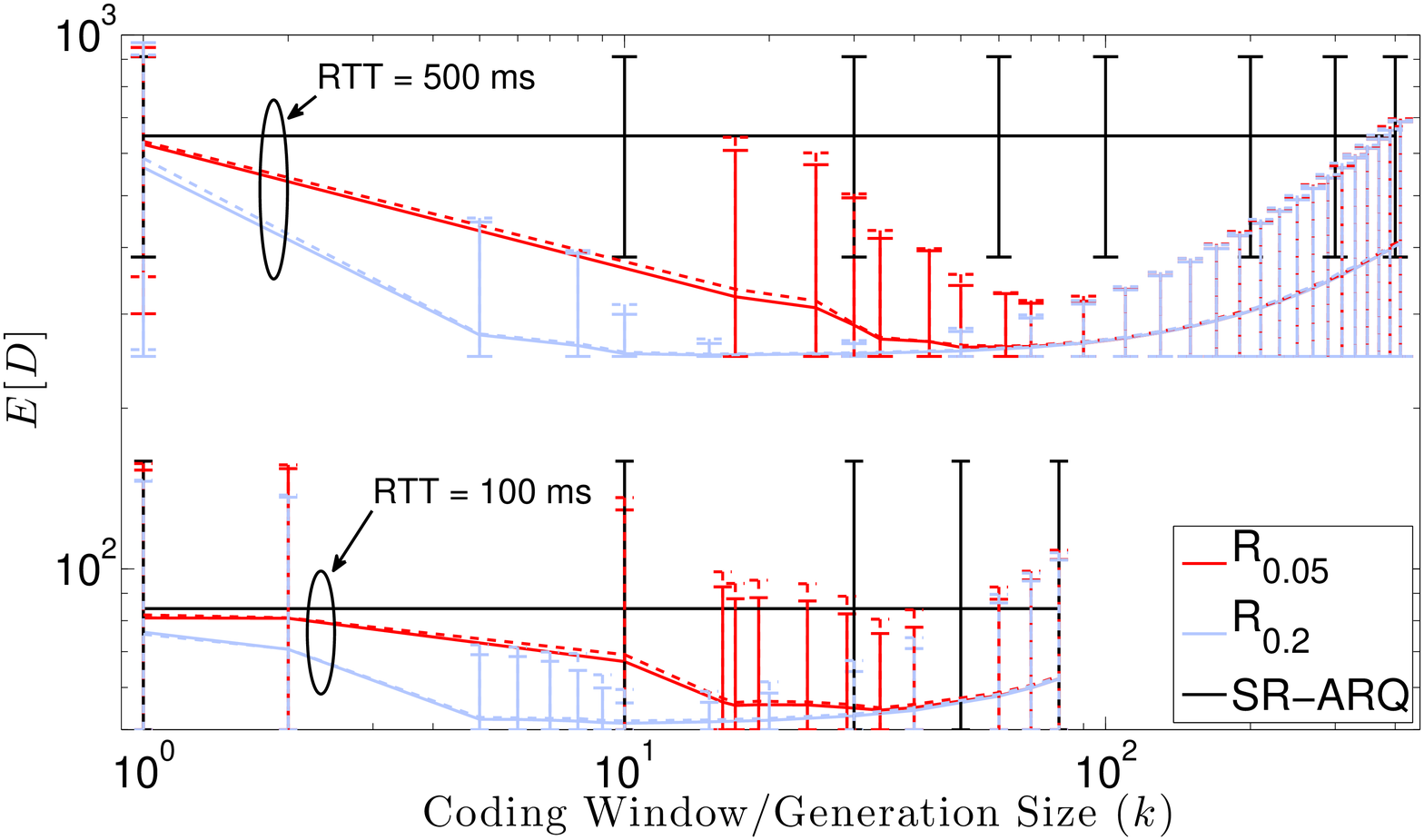}
\par\end{centering}

}\vspace{-10pt}
\par\end{centering}

\begin{centering}
\subfloat[$\epsilon=0.1$]{\begin{centering}
\includegraphics[width=1\columnwidth]{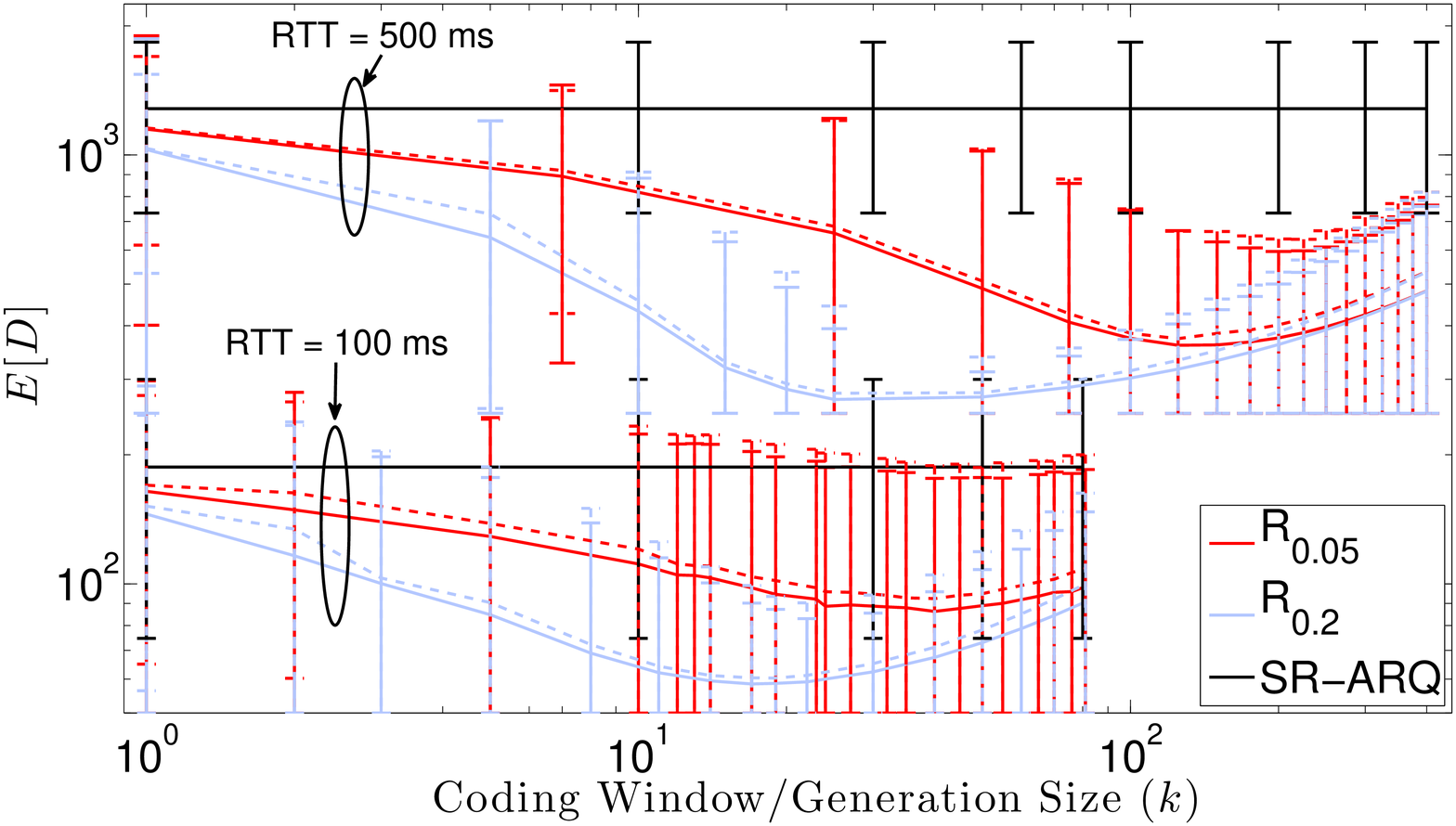}
\par\end{centering}

}
\par\end{centering}

\caption{The in-order delay for two erasure rates as a function of $k$ on
a 10 Mbps link. The error bars show $2\sigma_{D}$ above and below
the mean. The analytical and simulated results are represented using
solid and dotted lines respectively. Note the log scale of both the
x-axis and y-axis. \label{fig:Comparison-sim-analysis}}
\vspace{-15pt}
\end{figure}

Results for four different networks/links are shown in Figure \ref{fig:Comparison-sim-analysis}.
The simulation was developed in Matlab utilizing a model similar to
that presented in Section \ref{sec:Network-Model}, although several
of the assumptions are relaxed. The time it takes to retransmit coded
packets after feedback is received is taken into account. Furthermore,
the number of generations preventing delivery is not limited to a
single $BDP$ of packets, which increases the probability of head-of-line
blocking. Both of these relaxations effectively increases the delay
experienced by a packet. Finally, the figure shows the delay of an
idealized version of SR-ARQ where we assume infinite buffer sizes
and the delay is measured from the time a packet is first transmitted
until the time it is delivered in-order.

Figure \ref{fig:Comparison-sim-analysis} illustrates that adding
redundancy and/or choosing the correct coding window/generation size
can have major implications on the in-order delay. Not only does choosing
correctly reduce the delay, but can also reduce the jitter. However,
it is apparent when viewing $\mathbb{E}\left[D\right]$ as a function
of $k$ that the proper selection of $k$ for a given $R$ is critical
for minimizing $\mathbb{E}\left[D\right]$ and $\mathbb{E}\left[D^{2}\right]$.
In fact, Figure \ref{fig:Comparison-sim-analysis} indicates that
adding redundancy and choosing a moderately sized generation is needed
in most cases to ensure both are minimized.

Before proceeding, it is important to note that a certain level of
redundancy is needed to see benefits. Each curve shows results for
$R>\nicefrac{1}{1-\epsilon}$. For $R\leq\nicefrac{1}{1-\epsilon}$,
it is possible to see in-order delays and jitter worse than the idealized
ARQ scheme. Consider an example where a packet loss is observed near
the beginning of a generation that cannot be decoded after the first
transmission attempt. Since feedback is not sent/acted upon until
the end of the generation, the extra time waiting for the feedback
can induce larger delays than what would have occurred under a simple
ARQ scheme. We can reduce this time by reacting to feedback before
the end of a generation; but it is still extremely important to ensure
that the choice of $k$ and $R$ will decrease the probability of
a decoding failure and provide improved delay performance.

\begin{figure}
\begin{centering}
\includegraphics[width=1\columnwidth]{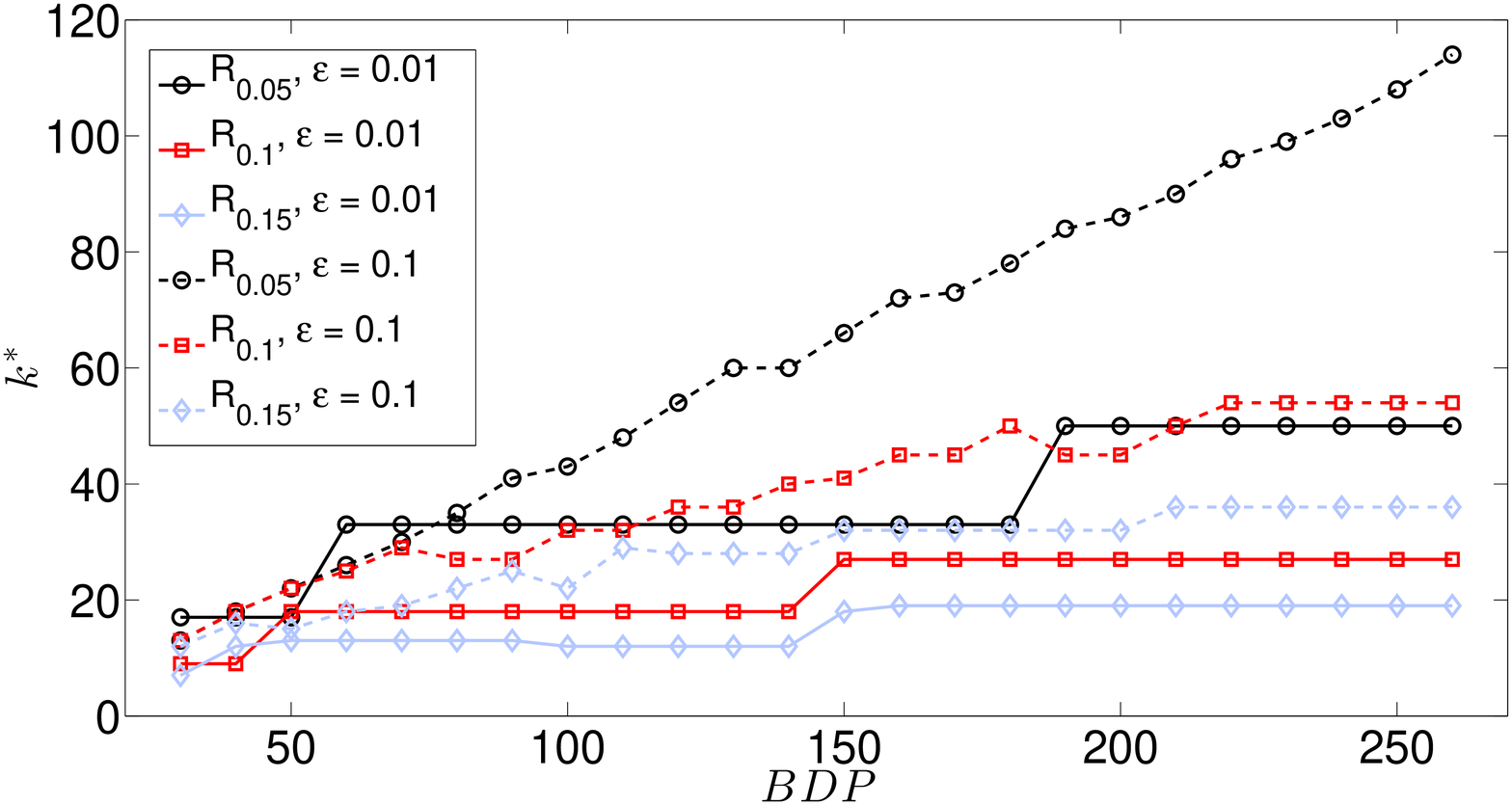}
\par\end{centering}

\caption{$k^{*}$ as a function of the $BDP$.\label{fig:MinK-BDP-tradeoff}}
\vspace{-15pt}
\end{figure}

The shape of the curves in the figure also indicate that there are
two major contributors to the in-order delay that need to be balanced.
Let $k^{*}$ be the generation size where $\mathbb{E}\left[D\right]$
is minimized for a given $\epsilon$ and $R$, i.e., 
\begin{equation}
k^{*}=\arg\min_{k}\mathbb{E}\left[D\right].\label{eq:min_k-1}
\end{equation}

To the left $k^{*}$, the delay is dominated by head-of-line blocking
and resequencing delay created by previous generations. To the right
of $k^{*}$, the delay is dominated by the time it takes to receive
enough $dofs$ to decode the generation. While there are gains in
efficiency for $k>k^{*}$, the benefits are negligible for most time-sensitive
applications. As a result, we show $k^{*}$ for a given $\epsilon$
and $R$ as a function of the $BDP$ in Figure \ref{fig:MinK-BDP-tradeoff}
and make three observations concerning this figure. First, the coding
window size $k^{*}$ increases with $\epsilon$, which is opposite
of what we would expect from a typical erasure code \cite{koetter_coding_2008}.
In the case of small $\epsilon$, it is better to try and quickly
correct only some of the packet losses occurring within a generation
using the initially transmitted coded packets while relying heavily
on feedback to overcome any decoding errors. In the case of large
$\epsilon$, a large generation size is better where the majority
of packet losses occurring within a generation are corrected using
the initially transmitted coded packets and feedback is relied upon
to help overcome the rare decoding error. Second, increasing $R$
decreases $k^{*}$. This due to the receiver's increased ability to
decode a generation without having to wait for retransmissions. Third,
$k^{*}$ is not very sensitive to the $BDP$ (in most cases) enabling
increased flexibility during system design and implementation.

\subsection{Rate-Delay Trade-Off}

\begin{figure}
\begin{centering}
\includegraphics[width=1\columnwidth]{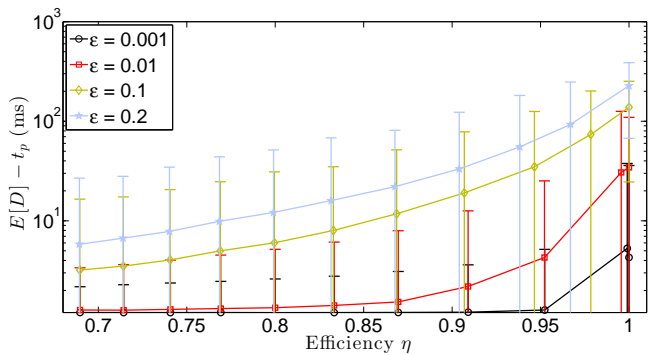}
\par\end{centering}

\caption{Rate-delay trade-off for a 10 Mbps link with a $RTT$ of 100 ms. The
error bars represent $2\sigma_{D}$ above and below the mean, and
the delay for ARQ is shown for $\eta=1$. Note the log scale of the
y-axis.\label{fig:Rate-delay-trade-off}}
\vspace{-17pt}
\end{figure}

While transport layer coding can help meet strict delay constraints,
the decreased delay comes at the cost of throughput, or efficiency.
Let $\mathbb{E}\left[D^{*}\right]$, $\sigma_{D}^{*}$, and $\eta^{*}$
be the expected in-order delay, the standard deviation, and the expected
efficiency respectively that corresponds to $k^{*}$ defined in eq.
(\ref{eq:min_k-1}). The rate-delay trade-off is shown by plotting
$\mathbb{E}\left[D^{*}\right]$ as a function of $\eta^{*}$ in Figure
\ref{fig:Rate-delay-trade-off}. The expected SR-ARQ delay (i.e.,
the data point for $\eta=1$) is also plotted for each packet erasure
rate as a reference.

The figure shows that an initial increase in $R$ (or a decrease in
$\eta$) has the biggest effect on $\mathbb{E}\left[D\right]$. In
fact, the majority of the decrease is observed at the cost of just
a few percent (2-5\%) of the available network capacity when $\epsilon$
is small. As $R$ is increased further, the primary benefit presents
itself as a reduction in the jitter (or $\mathbb{E}\left[D^{2}\right]$).
Furthermore, the figure shows that even for high packet erasure rates
(e.g., $20\%$), strict delay constraints can be met as long as the
user is willing to sacrifice throughput.

\subsection{Real-World Comparison\label{sub:Real-World-Comparison}}

\begin{figure}
\begin{centering}
\includegraphics[width=1\columnwidth]{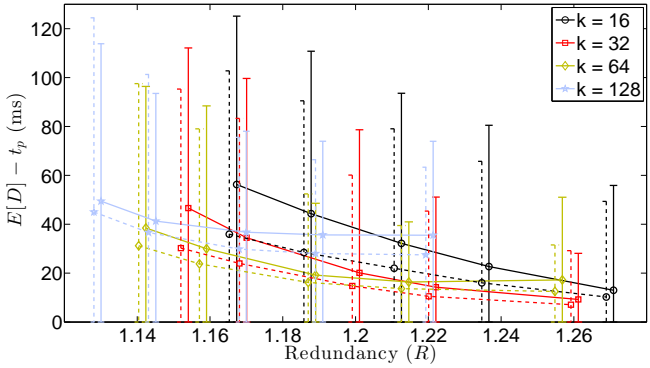}
\par\end{centering}

\caption{Experimental (solid lines) and analytical (dotted lines) results for
various $k$ over a 25 Mbps link with $RTT=60\text{ ms}$ and $\epsilon=0.1$.\label{fig:Experimental-in-order-delivery}}
\vspace{-20pt}
\end{figure}

We finally compare the analysis with experimentally obtained results
in Figure \ref{fig:Experimental-in-order-delivery} and show that
our analysis provides a reasonable approximation to real-world protocols.
The experiments were conducted using Coded TCP (CTCP) over an emulated
network similar to the one used in \cite{kim_congestion_2014} with
a rate of 25 Mbps and a $RTT$ of 60 ms. The only difference between
our setup and theirs was that we fixed CTCP's congestion control window
size ($cwnd$) to be equal to the $BDP$ of the network in order to
eliminate the affects of fluctuating $cwnd$ sizes.

There are several contributing factors for the differences between
the experimental and analytical results shown in the figure. First,
the analytical model approximates the algorithm used in CTCP. Where
we assume feedback is only acted upon at the end of a generation,
CTCP proactively acts upon feedback and does not wait until the end
of a generation to determine if retransmissions are required. CTCP's
standard deviation is less than the analytical standard deviation
as a result. Second, the experiments include additional processing
time needed to accomplish tasks such as coding and decoding, while
the analysis does not. Finally, the assumptions made in Sections \ref{sec:Network-Model}
and \ref{sec:Expected-In-Order-Delivery} effectively lower bounds
$\mathbb{E}\left[D\right]$ and $\mathbb{E}\left[D^{2}\right]$. Regardless,
the analysis does provide a fairly good estimate of the in-order delay
and can be used to help inform decisions regarding the appropriate
generation size to use for a given network/link.\vspace{-5pt}

\section{Conclusion\label{sec:Conclusion}}

In this paper, we addressed the use of transport layer coding to improve
application layer performance. A coding algorithm and an analysis
of the in-order delivery delay's first two moments were presented,
in addition to numerical results addressing when and how much redundancy
should be added to a packet stream to meet a user's delay constraints.
These results showed that the coding window size that minimizes the
expected in-order delay is largely insensitive to the $BDP$ of the
network for some cases. Finally, we compared our analysis with the
measured delay of an implemented transport protocol, CTCP. While our
analysis and the behavior of CTCP do not provide a one-to-one comparison,
we illustrated how our work can be used to help inform system decisions
when attempting to minimize delay.

\section*{Acknowledgments}

We would like to thank the authors of \cite{kim_congestion_2014}
for the use of their CTCP code. Without their help, we would not have
been able to collect the experimental results.

\bibliographystyle{ieeetr}
\bibliography{Improving_inorder_delivery_delay_using_transport_layer_coding}

\end{document}